\documentclass[letterpaper,number]{elsarticle}
\usepackage{amsmath}
\usepackage{amssymb}
\usepackage{graphicx}
\usepackage{color}

\begin{document}
\title{Nucleon Structure from Lattice QCD Using a Nearly Physical Pion Mass}
\author[mit,mainz]{J.~R.~Green}
\author[unm]{M.~Engelhardt}
\author[juelich]{S.~Krieg}
\author[mit]{J.~W.~Negele}
\author[mit]{A.~V.~Pochinsky}
\author[lbl,rbrc]{S.~N.~Syritsyn}\ead{ssyritsyn@quark.phy.bnl.gov}
\address[mit]{Massachusetts Institute of Technology, Cambridge, Massachusetts 02139, USA}
\address[mainz]{Institut f\"ur Kernphysik, Johannes Gutenberg-Universit\"at Mainz, 
      D-55099 Mainz, Germany}
\address[unm]{Department of Physics, New Mexico State University, Las Cruces, New Mexico 88003, USA}
\address[juelich]{Bergische Universit\"at Wuppertal, D-42119 Wuppertal, Germany and
IAS, J\"ulich Supercomputing Centre, Forschungszentrum J\"ulich, D-52425 J\"ulich, Germany}
\address[lbl]{Lawrence Berkeley National Laboratory, Berkeley, California 94720, USA}
\address[rbrc]{RIKEN/BNL Research Center, Brookhaven National Laboratory, Upton, NY 11973 USA}

\date{\today}

\begin{abstract}
We report the first Lattice QCD calculation using the almost physical pion
mass $m_\pi=149\text{ MeV}$ that agrees with experiment for four fundamental
isovector observables characterizing the gross structure of the nucleon:
the Dirac and Pauli radii, the magnetic moment, and the quark momentum
fraction.  The key to this success is the combination of using a nearly physical 
pion mass and excluding the contributions of excited states.  
An analogous calculation of the nucleon axial charge governing beta
decay has inconsistencies indicating a source of bias at low pion masses
not present for the other observables and yields a result that disagrees 
with experiment.
\end{abstract}
\begin{keyword}
lattice QCD, nucleon structure, form factors
\PACS 12.38.Gc \sep 13.60.Fz \sep 14.20.Dh
\end{keyword}

\maketitle

\section{Introduction}

Lattice QCD is the only known rigorous framework for ab-initio calculation 
of the structure of protons and neutrons with controllable uncertainties.
It can provide quantitative answers to both fundamental questions 
such as the quark and gluon composition of the nucleon spin 
and phenomenological questions such as the sensitivity of modern detectors 
to physics beyond the Standard Model (BSM), to fundamental symmetry violations, 
and to hypothetical dark matter 
particles~\cite{Ellis:2009ai,Bhattacharya:2011qm,Green:2012ej}.
It also has the potential to help resolve discrepancies between inconsistent
experimental results.
However, with current computer resources, its predictive power is limited by 
uncertainties arising from heavier than physical quark masses, 
finite lattice spacing and volume, incomplete removal of excited states, 
and omission of disconnected contractions. Progress in lattice calculations
of nucleon structure hinges on identifying and removing the most significant
among these uncertainties.

Significant effort has been focused on lattice calculations 
of the isovector  Dirac and Pauli radii
$(r_{1,2}^2)^v$, anomalous magnetic moment $\kappa_v$, axial charge $g_A$, 
and quark momentum fraction
$\langle x\rangle_{u-d}$:
\begin{gather}
\label{eq:vector_ff_def}
\begin{gathered}
  \langle p^\prime | \bar{q}\gamma^\mu q |p\rangle 
    = \bar{u}_{p^\prime}\big[F^q_1(Q^2)\gamma^\mu 
        + F^q_2(Q^2)\frac{i\sigma^{\mu\nu}q_\nu}{2M}\big] u_p, \\
  F_{1,2}^q(Q^2) \underset{Q^2\to0}{\approx}
    F^q_{1,2}(0)\big(1 - \frac16(r_{1,2}^2)^q Q^2+ {\mathcal O}(Q^4)\big), \\
  \kappa_v = F_2^q(0)\,,
\end{gathered}
\\
\label{eq:momfrac_def}
\langle p | \bar{q}\gamma_{\{\mu} {\overset{\leftrightarrow}{D}}_{\nu\}}q | p\rangle 
  = \langle x\rangle_q \;\bar{u}_p \gamma_{\{\mu} p_{\nu\}} u_p ,
\\
\label{eq:gA_def}
\langle p | \bar{q}\gamma^\mu\gamma^5 q| p\rangle 
  = g_A\;\bar{u}_p \gamma^\mu\gamma^5 u_p\,,
\end{gather}
where $Q^2=-q^2 = -(p^\prime-p)^2$ and $u_p$, $u_{p^\prime}$ are nucleon
spinors, and we note that computationally expensive disconnected contractions
cancel in isovector observables.

Until recently, although some success has been
achieved~\cite{Edwards:2005ym,Hagler:2007xi,Yamazaki:2009zq,Alexandrou:2011db,
Collins:2011mk,Capitani:2012gj}, lattice results relied heavily on large
extrapolations using Chiral Perturbation Theory (ChPT) yielding potentially
uncontrollable corrections.
This is particularly problematic for some observables, e.g., $(r_{1,2}^2)^v$
and $\langle x\rangle_{u-d}$, for which ChPT predicts rapid change towards the
chiral regime.
For example, in typical lattice calculations with pion masses $\gtrsim250\text{
MeV}$, prior to extrapolation to $m_\pi^\text{phys}\approx135\text{ MeV}$,
$(r_1^2)^{v}$ is underestimated by
$\approx50\%$~\cite{Syritsyn:2009mx,Yamazaki:2009zq,Alexandrou:2011db,Collins:2011mk},
$\langle x\rangle_{u-d}$ overestimated by
$30-60\%$~\cite{Hagler:2007xi,Alexandrou:2011nr,Bali:2012av}, and $g_A$
underestimated by
$\approx10\%$~\cite{Edwards:2005ym,Yamazaki:2008py,Alexandrou:2010hf}, compared
with experiment.
These glaring discrepancies and the dependence on large extrapolations 
clearly indicate the need for calculations near the physical pion mass.
Moreover, it has recently been found that excited-state effects become worse 
with decreasing pion mass~\cite{Green:2011fg}, 
and their careful analysis is required before even attempting extrapolations 
in the pion mass towards the physical point using ChPT.

In this paper, we report the first Lattice QCD calculation of five nucleon
structure observables  using pion masses as light as $m_\pi=149\text{ MeV}$. 
This is so close to the physical value that chiral extrapolation of these
observables yields changes between the lowest pion mass point, $m_{\pi } =
149\text{ MeV}$, and the physical point,  $m_\pi^\text{phys} \approx 135\text{
MeV}$, that are less than or comparable to the
statistical uncertainty of the $m_{\pi } = 149\text{ MeV}$ data.
For each ensemble, we remove excited-state contaminations by varying the
source-sink separation in the range $\approx0.9 \ldots 1.4\text{ fm}$ and apply
the summation method \cite{Mathur:1999uf} to extract the ground state matrix
elements. 
We observe remarkable agreement with experiment, well within the statistical
uncertainties, for the isovector Dirac and Pauli radii, the anomalous magnetic
moment, and the quark momentum fraction, all computed with the same
methodology. 
However, as discussed below, there are inconsistencies in the results for the
axial charge, $g_A $, calculated at the  lowest two pion masses that do not
arise for the other observables, that make the results suspect, and that indeed
lead to disagreement with experiment. 
There appears to be a source of bias significantly affecting the evaluation of
the axial charge at low pion masses, leading to a qualitative difference in
behavior between $g_A $ and the other observables.

\section{Lattice Results}
We perform calculations using ten ensembles of Lattice QCD gauge fields
generated with ${\mathcal O}(a^2)$-improved Symanzik gauge action and
tree-level clover-improved Wilson fermion action using 2-HEX stout gauge
links~\cite{Durr:2010aw}. Two light $u$ and $d$ (with $m_u=m_d$) and one
heavier strange ($m_s\gg m_{u,d}$, tuned to be close to physical) quark
flavors are simulated fully dynamically, while effects of heavier quarks
are neglected. In addition to varying the pion mass in the range
$149\ldots 357\text{ MeV}$, we include different spatial volumes, time
extents of the lattice, and one ensemble with a smaller lattice spacing
in order to estimate the size of the corresponding systematic effects;
see Tab.~\ref{tab:lat_ens}. Nucleon matrix elements are extracted from
nucleon 3-point functions that are computed with the standard sequential
source method. Nucleon field operators are optimized to overlap as much
as possible with the single-nucleon ground state at rest by tuning the
spatial width of Gaussian smeared quark sources.

In order to discriminate between the ground and excited-state matrix
elements on a Euclidean lattice, we vary the timelike distance $\Delta t$
between nucleon sources and sinks in the 3-point functions.
With increasing $\Delta t$, excited states in 3-point functions are suppressed 
as $\sim e^{-\Delta E \Delta t/2}$ and disappear in the $\Delta t\to\infty$
limit, where $\Delta E$ is the (potentially $m_\pi$-dependent) energy gap
to the closest contributing state. However, using a large source-sink
separation is impractical, since statistical noise grows rapidly with
$\Delta t$. Instead, we combine calculations with three values of
$\Delta t\approx0.9$, $1.2$, $1.4\text{ fm}$ using the summation 
method~\cite{Mathur:1999uf}, which benefits from improved asymptotic
behavior~\cite{Capitani:2010sg,Bulava:2010ej} and which we find the most
reliable and robust method with the presently existing statistics.
In a separate publication~\cite{bmw-ff-inprep} we report detailed studies
and comparisons of different methods, 
including also the generalized pencil-of-function (GPoF) method~\cite{Aubin:2010jc}.

We present and discuss our results for all the observables 
in Figs.~\ref{fig:r1v2}--\ref{fig:gA-2p} below.
To emphasize the importance of controlling excited states, in the same figures 
we also show the standard plateau method results at given $\Delta t$
(open symbols) for the three lightest $m_\pi $. As this separation is
increased, the data points consistently approach our final results, 
while their uncertainties increase as expected. As illustrated by the
data, the effect of the excited states may be very dramatic, especially
for $(r_1^2)^v$, $\kappa^v(r_2^2)^v$ and $\langle x\rangle_{u-d}$.

\begin{table}[ht]
  \centering
  \caption{\label{tab:lat_ens}
    Lattice QCD ensembles. The strange quark mass $m_s$ is tuned to be close to physical.
  }
  \begin{tabular}{lc|ccc|rr}
    \hline\hline
    $a$ [fm] & $L_s^3\times L_t$ & 
      $m_\pi\text{ [MeV]}$ & $m_\pi L_s$ & $m_\pi L_t$ &
      \# confs & \# meas 
    \\
    \hline
    0.116 & $48^3\times 48$ & 149(1) & 4.20 & 4.20 & 646 & 7752 \\
    0.116 & $32^3\times 48$ & 202(1) & 3.80 & 5.70 & 457 & 5484 \\
    0.116 & $32^3\times 96$ & 253(1) & 4.78 & 14.3 & 202 & 2424 \\
    0.116 & $32^3\times 48$ & 254(1) & 4.78 & 7.17 & 420 & 5040 \\
    0.116 & $24^3\times 48$ & 254(1) & 3.58 & 7.17 & 1019 & 24456 \\
    0.116 & $24^3\times 24$ & 252(2) & 3.57 & 3.57 & 3999 & 23994 \\
    0.116 & $24^3\times 48$ & 303(2) & 4.28 & 8.55 & 128 & 768 \\
    0.090 & $32^3\times 64$ & 317(2) & 4.64 & 9.28 & 103 & 824 \\
    0.116 & $24^3\times 24$ & 351(2) & 4.97 & 4.97 & 420 & 2520 \\
    0.116 & $24^3\times 48$ & 356(2) & 5.04 & 10.1 & 127 & 762 \\
  \hline\hline
  \end{tabular}
\end{table}

\begin{table}[ht]
  \centering
  \caption{\label{tab:fits}Extrapolations to the isospin-limit 
    $m_\pi^\text{phys}=134.8\text{ MeV}$~\cite{Colangelo:2010et}
    using data points $149\le m_\pi \le 254\text{ MeV}$. 
    The footnotes describe details of the ChPT formulas
    and fixed parameters used in the fits.
    The right column shows the deviation from experiment, 
    relative to the lattice uncertainty.}
  \begin{tabular}{cl|lc|lcc}
  \hline\hline
  $X$ &  & $X^\text{lat}$ & $\chi^2/\text{dof}$ & 
    $X^\text{exp}$ & & $\frac{X^\text{lat} - X^\text{exp}}{\delta X^\text{lat}}$ \\
  \hline 
  $(r_1^2)^v [\text{fm}^2]$ & 
  ${}^a$ &
    $0.577(29)$ & 
    $0.8/2$ & 
    $0.579(3)$ & \cite{Pohl:2010zza}${}^f$ & 
    $-0.07$ \\
  $\kappa_v$ &
      ${}^{ab}$ & 
    $3.69(38)$ & 
    $1.3/1$ & 
    $3.706$ & \cite{Beringer:1900zz} &
    $-0.04$ \\
  $\kappa_v(r_2^2)^v [\text{fm}^2]$ &
      ${}^{ac}$ &
    $2.46(25)$ & 
    $1.2/2$ & 
    $2.47(7)$ & \cite{Beringer:1900zz} &
    $-0.04$ \\
  $\langle x\rangle_{u-d}$ &
      ${}^{d}$ &
    $0.140(21)$ & 
    $0.9/1$ & 
    $0.155(5)$ & \cite{durham-phenom-PDF} &
    $-0.71$ \\
  $g_A$ &
    ${}^{e}$ &
    $0.97(8)$ & 
    $0.13/1$ & 
    $1.2701(25)$ & \cite{Beringer:1900zz} &
    $-3.75$ \\
  \hline\hline
  \end{tabular}\\
  \flushleft
  {\footnotesize ${}^a$
    SSE ${\mathcal O}(\epsilon^3)$~\cite{Bernard:1998gv} with fixed 
    $F_\pi^0=86.2\text{ MeV}$~\cite{Colangelo:2003hf}, 
    $\Delta=293\text{ MeV}$~\cite{Syritsyn:2009mx}, 
    $g_A^0=1.26$~\cite{Hemmert:2003cb,Edwards:2005ym,Khan:2006de,Procura:2006gq}, 
    $c_A=1.5$~\cite{Syritsyn:2009mx}.} \\
  {\footnotesize ${}^b$ Includes $(N\Delta)_{M1}$
    transition with $c_V=-2.5\text{ GeV}^{-1}$~\cite{Hemmert:2002uh}.}\\
  {\footnotesize ${}^c$ Includes higher-order ``core'' term~\cite{Gockeler:2003ay}.}\\
  {\footnotesize ${}^d$ BChPT ${\mathcal O(p^2)}$~\cite{Dorati:2007bk} with fixed
    $F_\pi^0=86.2\text{ MeV}$~\cite{Colangelo:2003hf}, 
    $g_A^0=1.26$~\cite{Hemmert:2003cb,Edwards:2005ym,Khan:2006de,Procura:2006gq}, 
    $M_N^0=0.873$~\cite{Syritsyn:2009mx}, 
    $\Delta a_{20}^v=0.165$~\cite{Edwards:2006qx}.}\\
  {\footnotesize ${}^e$ SSE ${\mathcal O}(\epsilon^3)$~\cite{Hemmert:2003cb} with fixed 
    $F_\pi^0=86.2\text{ MeV}$~\cite{Colangelo:2003hf},
    $\Delta=293\text{ MeV}$~\cite{Syritsyn:2009mx},
    $c_A=1.5$~\cite{Syritsyn:2009mx}, $g_1=2.5$~\cite{Bratt:2010jn}.}\\
  {\footnotesize ${}^f$ From $r_E^p$~\cite{Pohl:2010zza} and $(r_E^2)^n$~\cite{Beringer:1900zz}.
    Using Ref.~\cite{Beringer:1900zz} for both $(r_E^2)^{p,n}$ results in 
    $(r_1^2)^v=0.640(9)$ (higher exp. point in Fig.~\ref{fig:r1v2}).}
\end{table}

Our lightest pion mass value $m_\pi=149(1)\text{ MeV}$ is only $\approx10\%$
higher than the physical pion mass in the isospin limit
$m_\pi\approx134.8\text{ MeV}$~\cite{Colangelo:2010et}.
To make  the small extrapolation of our data in $m_\pi$ to the physical point,
we add the $202(1)\text{ MeV}$ and the large-volume $254(1)\text{ MeV}$
ensembles  to  minimize dependence on the functional form and volume  and
perform  ChPT fits using standard parameters from ChPT phenomenology.
Table~\ref{tab:fits} gives details of the fits, results and comparisons with
experiment. 
Note that  due to the short extrapolation distance, the change in the values of
the chiral fits between $m_\pi=149\text{ MeV}$ and $m_\pi=135\text{ MeV}$ is
smaller than the statistical uncertainties of the $m_\pi=149\text{ MeV}$ data
in all cases  except for $(r_1^2 )^v $, for which the two are comparable. 
Although they are not shown, again because of the short extrapolation distance
linear fits to the data yield extrapolated values statistically consistent with
the chiral  fits.  Hence, the validity of our final extrapolated results is not
dependent on the convergence of ChPT and in contrast to all previous
calculations, the dominant uncertainties in the final results are the
statistical uncertainties in the  lowest mass data points rather than the
chiral extrapolations.
To explore the consistency of our data with the low-energy theory, we repeat
the chiral fits using the full available range of $m_\pi=149\ldots357\text{
MeV}$ and show the error bands for both ranges in
Figs.~\ref{fig:r1v2}--\ref{fig:gA-2p}.  
For $(r_1^2)^v$, $(r_2^2)^v$, $\kappa_v$, and $\langle x\rangle_{u-d}$ results
from $m_\pi\lesssim350\text{ MeV}$ and $m_\pi\lesssim250\text{ MeV}$ fits agree
within statistical uncertainty whereas for $g_A$, both the error bands and the
extrapolated result are inconsistent,  providing one indication of an
unphysical bias in the calculation of $g_A$ to be discussed below.

\begin{figure}[ht]
  \centering
  \includegraphics[width=.68\textwidth]{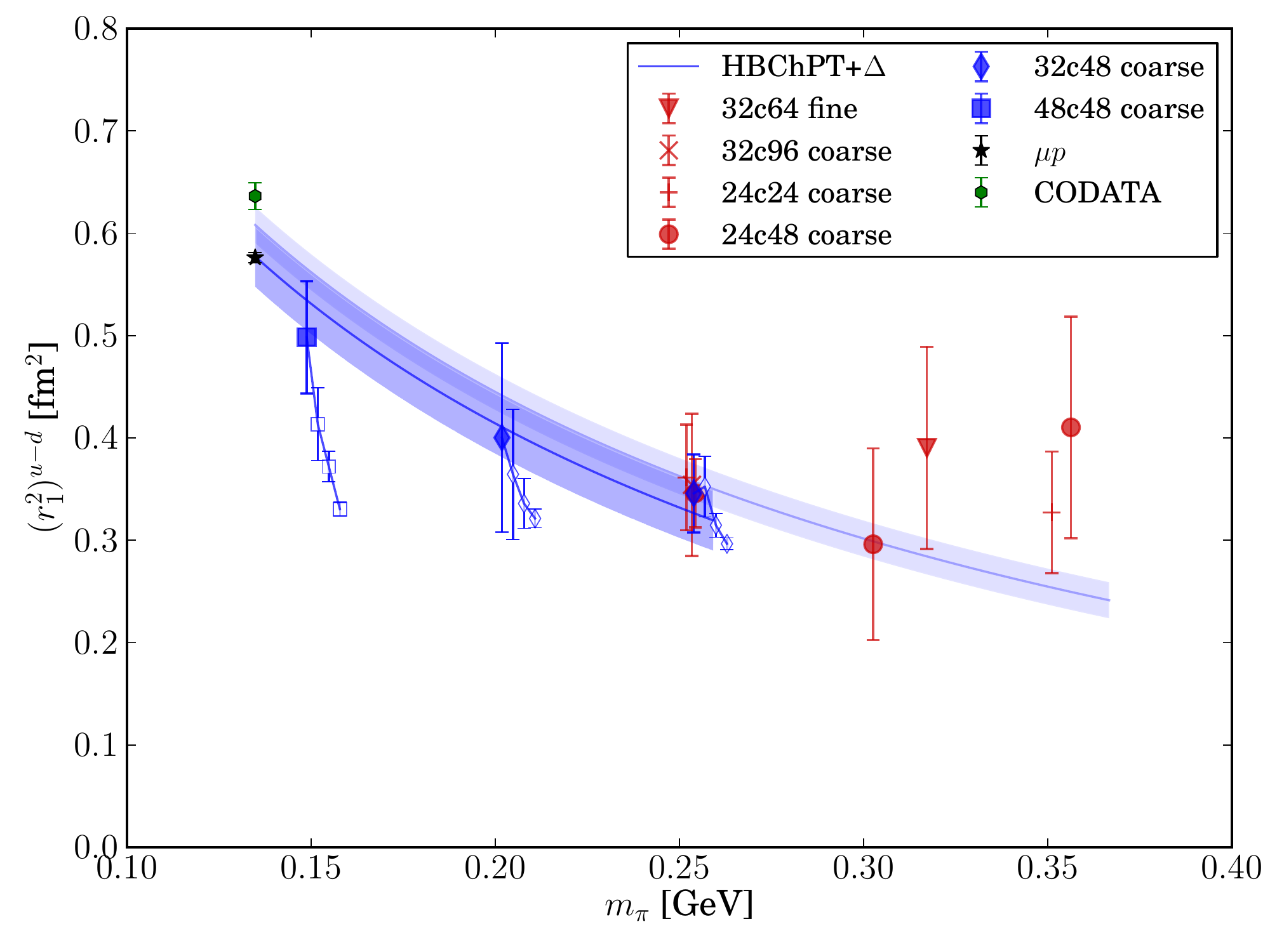}
  \caption{\label{fig:r1v2}
    Isovector Dirac radius $(r_1^2)^v$. 
    Fits to the solid square and diamond points are described in Tab.~\ref{tab:fits}, 
    and the same fits applied to the full set of solid points are shown for comparison.
    One experimental point is the CODATA recommended value~\cite{Mohr:2012tt} 
    and the other is from the $\mu p$ Lamb shift~\cite{Pohl:2010zza}.
    The series of open symbols show data before the removal of excited states, 
    with fixed source-sink separation $\Delta t$ increasing from right to left.
    Their error bars reflect only statistical uncertainties, which grow with $\Delta t$.
    }
\end{figure}
\begin{figure}[ht]
  \centering
  \includegraphics[width=.68\textwidth]{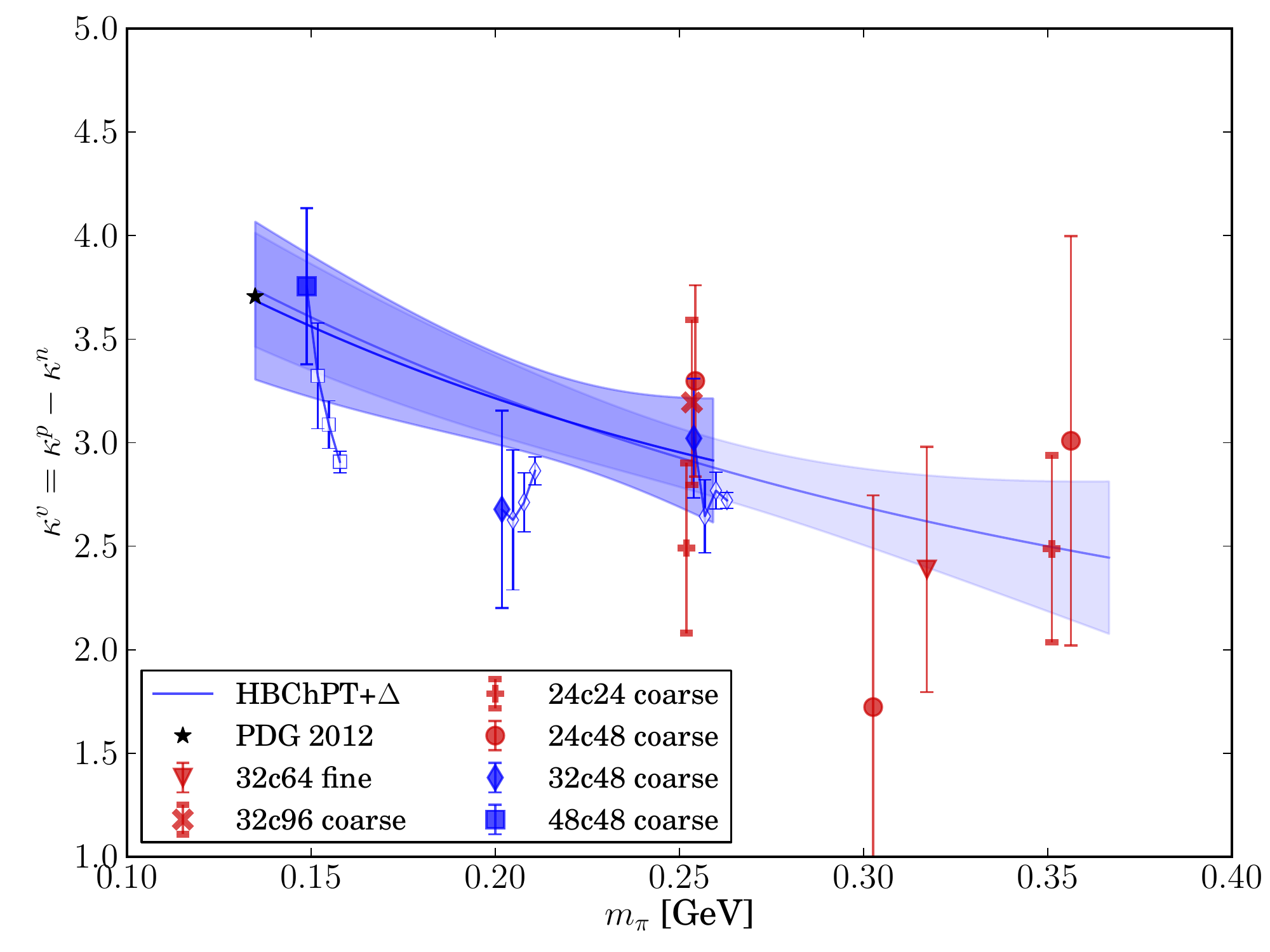}
  \caption{\label{fig:kvnorm}
    Isovector anomalous magnetic moment $\kappa^v$.
    See caption of Fig.~\ref{fig:r1v2}.}
\end{figure}
\begin{figure}[ht]
  \centering
  \includegraphics[width=.68\textwidth]{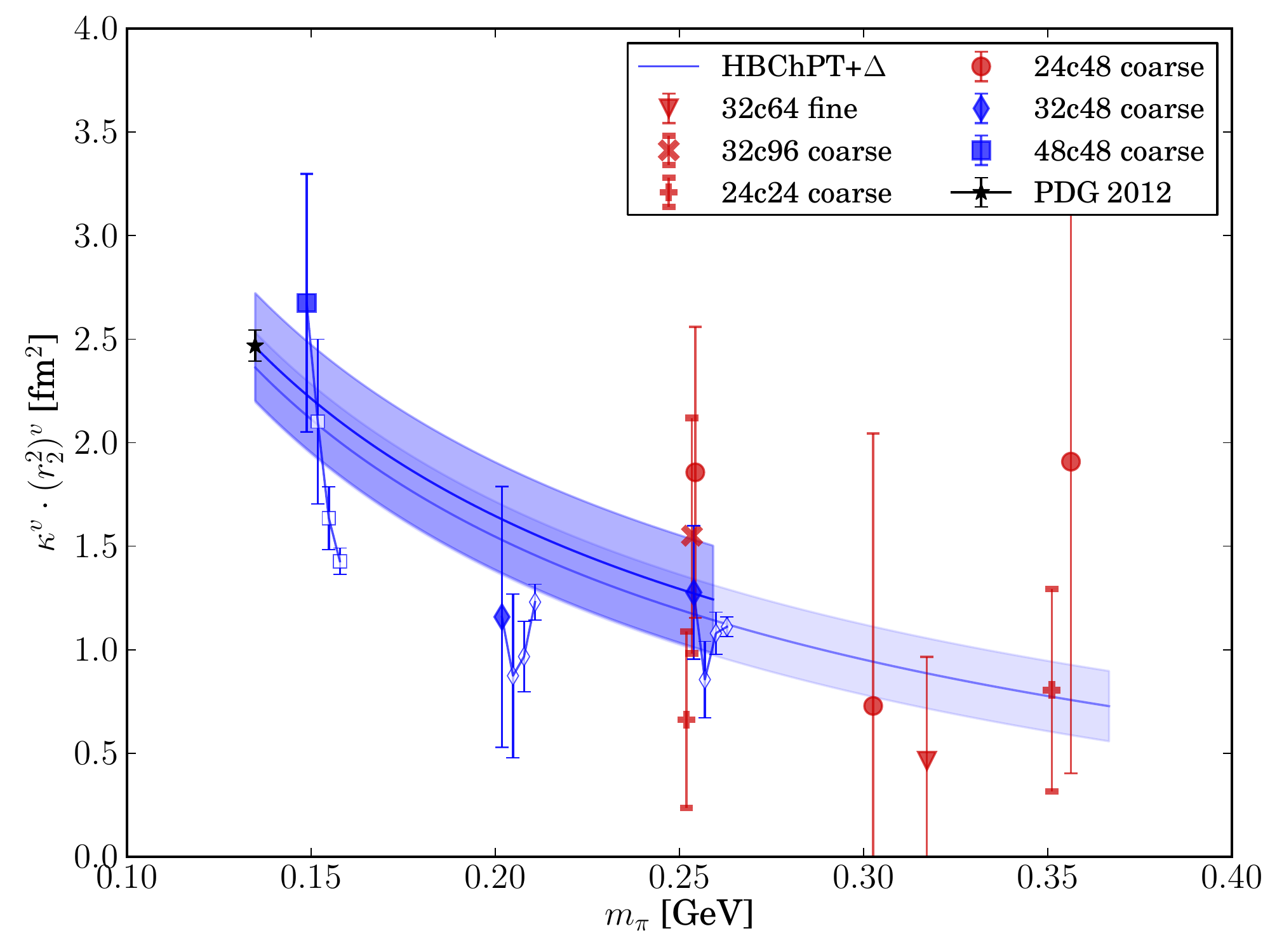}
  \caption{\label{fig:r2v2_kv}
    Isovector Pauli radius $\kappa^v(r_2^2)^v$.
    See caption of Fig.~\ref{fig:r1v2}.}
\end{figure}
\begin{figure}[ht]
  \centering
  \includegraphics[width=.68\textwidth]{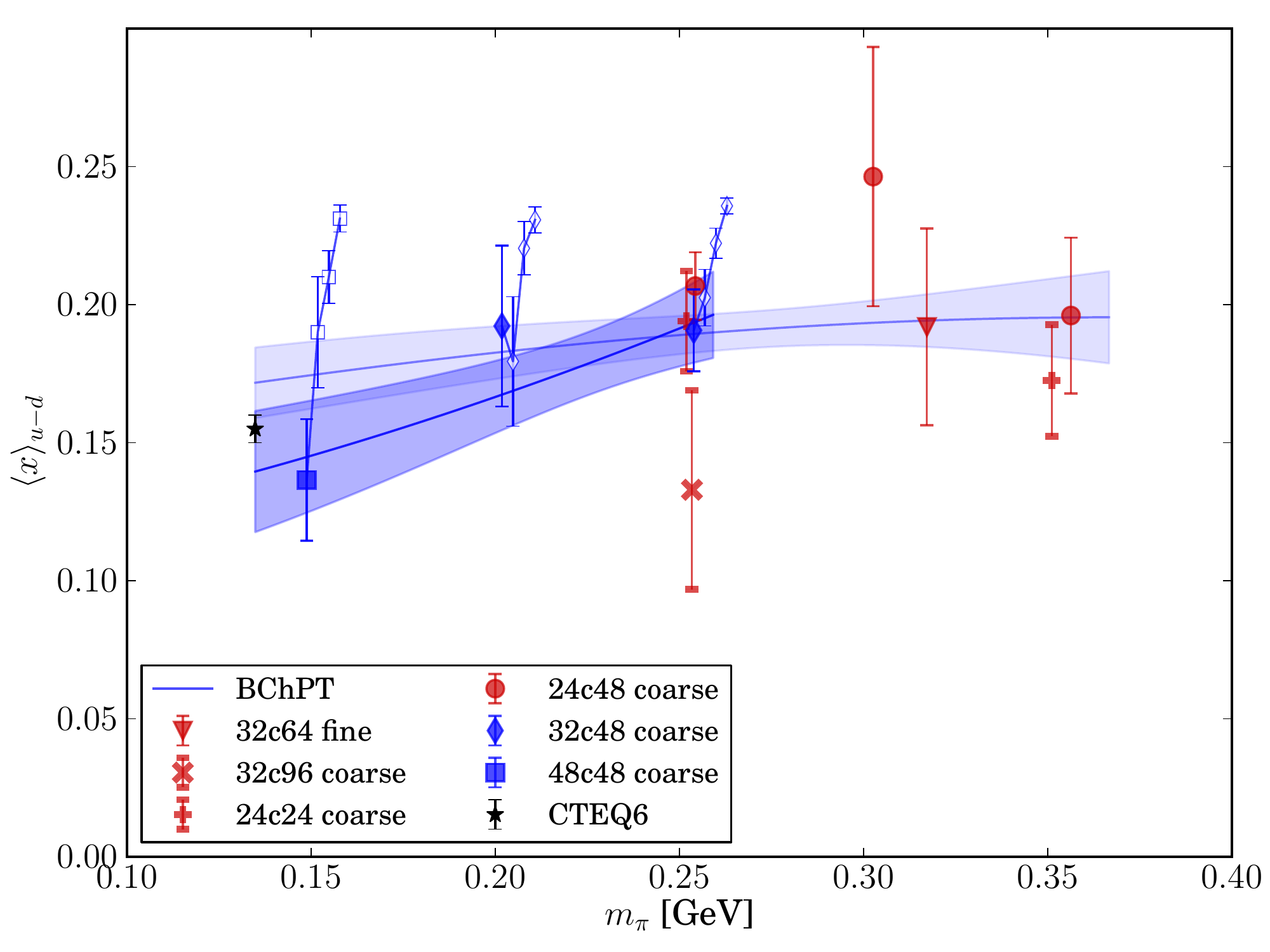}
  \caption{\label{fig:momfrac}
    Isovector quark momentum fraction $\langle x\rangle_{u-d}$.
    See caption of Fig.~\ref{fig:r1v2}.}
\end{figure}
\begin{figure}[ht]
  \centering
  \includegraphics[width=.68\textwidth]{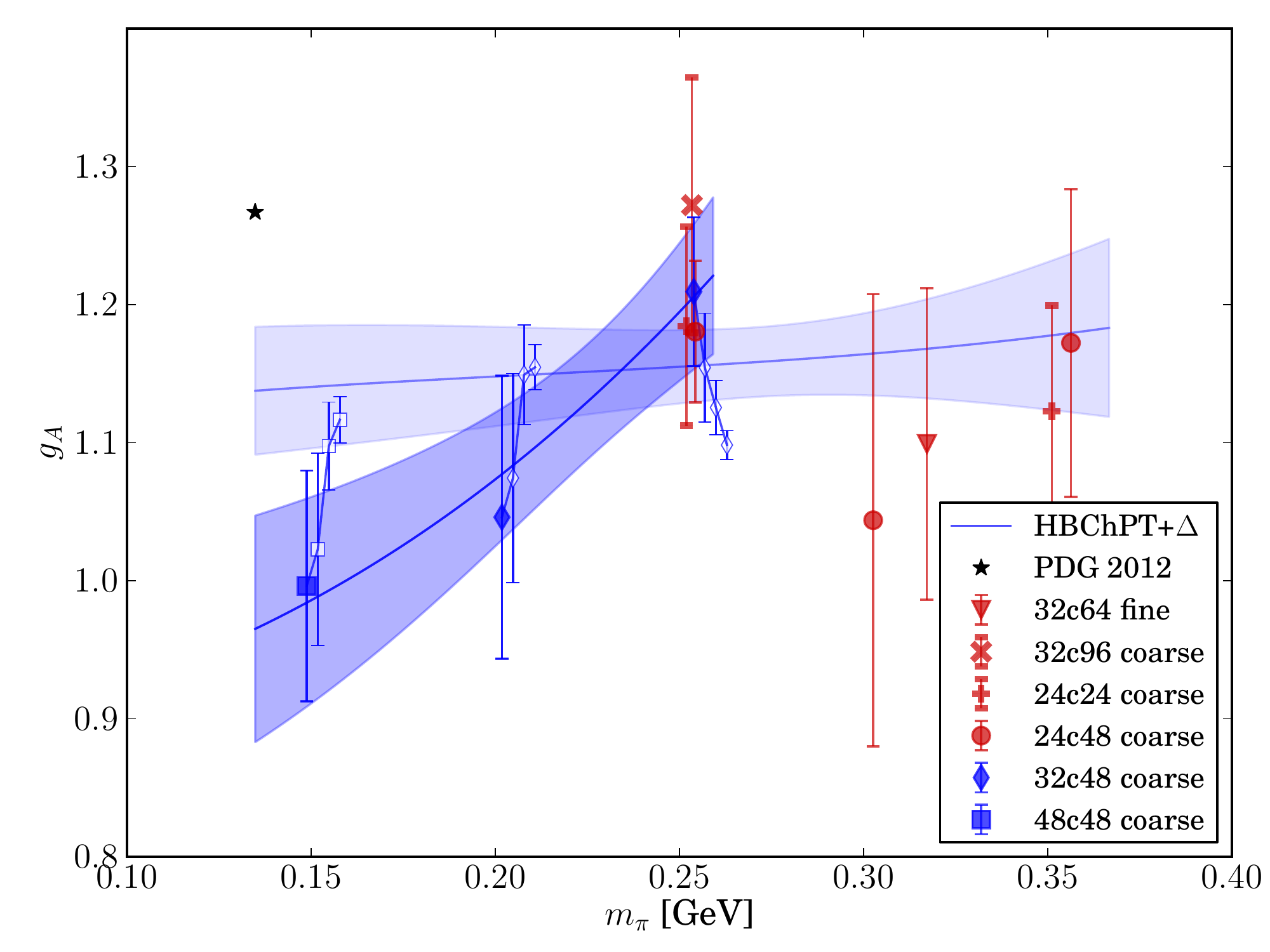}
  \caption{\label{fig:gA-2p}
    Axial charge $g_A$. 
    See caption of Fig.~\ref{fig:r1v2}.
       }
\end{figure}

To explore nucleon electromagnetic structure, we compute matrix elements of the
quark-nonsinglet vector current $\langle p^\prime | \big[\bar{u}\gamma_\mu u -
\bar{d}\gamma_\mu d\big] | p\rangle$ between polarized proton ($uud$) states
with different momenta $p$, $p^\prime$.  We extract isovector Dirac and Pauli
form factors $F_{1,2}^v(Q^2)$ from these matrix elements, and then the
``radii'' $(r_{1,2}^2) = -\frac6{F_{1,2}}\frac{dF_{1,2}}{dQ^2}\Big|_{Q^2=0}$
using dipole fits $F_{1,2}(Q^2)\sim\frac{F_{1,2}(0)}{(1+Q^2/M_D^2)^2}$ in the
range $0 \le Q^2\le 0.5\text{ GeV}^2$.

The Dirac radius $(r_1^2)^v$ is shown in Fig.~\ref{fig:r1v2}. 
We compare it to the experimental value $(r_1^2)^v=(r_1^2)^p - (r_1^2)^n$,
where $(r_1^2)^{p,n} = (r_E^2)^{p,n} -\frac{3\kappa^{p,n}}{2 M_{p,n}^2}$, with
the error bar dominated by the uncertainty in $(r_E^2)^p$, the proton electric
charge radius.
We show two experimental values for $(r_1^2)^v$ in Fig.~\ref{fig:r1v2}, which
correspond to two inconsistent values for $(r_E^2)^p$: the CODATA recommended
value~\cite{Mohr:2012tt} (also cited by PDG~\cite{Beringer:1900zz})
based on hydrogen spectroscopy and $e - p$ scattering
and the recent and controversial result from measurement of the $\mu - p$ Lamb
shift~\cite{Pohl:2010zza}\footnote{
  It is notable that the new experimental value from the $\mu-p$ Lamb shift is in 
  better agreement with 
  phenomenological fits based on dispersion relations~\cite{Lorenz:2012tm}.
}.
Relative to the lattice uncertainty, the extrapolated value deviates from the
$\mu p$ Lamb shift value by $-0.07\sigma_\text{lat}$ and from the PDG value by
$-2.17\sigma_\text{lat}$, where $\sigma_\text{lat}$ is the uncertainty of the
present calculation.

In a similar fashion, we extract the isovector anomalous magnetic moment
$\kappa^v = \kappa^p-\kappa^n$ and the Pauli radius $(r_2^2)^v =
(\kappa^p(r_2^2)^p - \kappa^n(r_2^2)^n)/(\kappa^p-\kappa^n)$ from the Pauli
form factor $F_2(Q^2)$.
The results are less precise than $(r_1^2)^v$ because the forward
values $F_2(0)$ and $\frac{dF_2(0)}{dQ^2}$ must be extrapolated, and we use
the dipole form $F_2^v(Q^2)\sim \frac{\kappa^v}{(1+Q^2/M_{D2}^2)^2}$. 
Since the minimal value $Q^2>0$ scales as $Q^2_\text{min}\sim\frac1{L_s^2}$,
the $Q^2$ fit is less precise in smaller  spatial volumes.
This explains the significant increase of error bars in Figs.~\ref{fig:kvnorm},
\ref{fig:r2v2_kv}  going from  $32^3$ to $24^3$   lattices (e.g., at
$m_\pi\approx250\text{ MeV}$), compared to the corresponding error bars of
$(r_1^2)^v$ in Fig.~\ref{fig:r1v2}. 
Fortunately, our calculation at the lightest pion mass uses the largest
$V_s\approx(5.6\text{ fm})^3$  yielding the smallest
$Q_\text{min}^2\approx0.05\text{ GeV}^2$  and is the least affected by this
problem.


We compute the isovector quark momentum fraction\footnote{
  $\langle x\rangle_{u-d}=\langle x\rangle_u-\langle x\rangle_d$ is understood
  as the momentum fraction carried by quarks \emph{and} antiquarks, i.e. $\langle
  x\rangle_q =\int_0^1 dx \; x \; \big( f_q(x) + f_{\bar q}(x) \big)$, where
  $f_{q(\bar{q})}$ is a parton distribution function (PDF).}
$\langle x\rangle_{u-d}$
from the forward matrix element of the operator in Eq.~(\ref{eq:momfrac_def})
and renormalize the lattice value to the standard $\overline{\text{MS}}(2\text{
GeV})$ scheme using the $RI^\prime / MOM$ method~\cite{Martinelli:1994ty}.
In Fig.~\ref{fig:momfrac} we show our results together with the CTEQ6
value~\cite{durham-phenom-PDF}.
Achieving agreement between Lattice QCD and phenomenology for $\langle
x\rangle_{u-d}$ is one of the most important accomplishments of this paper:
previous lattice calculations~\cite{Bratt:2010jn,Aoki:2010xg,Bali:2012av}
consistently overestimated\footnote{
  There is a hint in a recent lattice calculation in Ref.~\cite{Bali:2012av} that 
  $\langle x\rangle_{u-d}$ decreases and becomes closer to phenomenology
  as the pion mass approaches the physical value.}
the phenomenological value. 
The combination of our use of a nearly physical pion mass and  removal of
excited state contamination has eliminated the discrepancy.
The presence of substantial excited-state contaminations in $\langle x\rangle_{u-d}$ 
was first hinted at in Ref.~\cite{Lin:2008uz} for $m_\pi = 493\text{ MeV}$, 
later substantiated by an early report from the present study~\cite{Green:2011fg}
and at $m_\pi = 373\text{ MeV}$~\cite{Dinter:2011sg}, 
and recently further confirmed at 
$m_\pi \geq 195\text{ MeV}$~\cite{Jager:2013kha}.

In contrast to the observables discussed above, the computed value of the
nucleon axial charge $g_A$ disagrees significantly  with
$g_A^\text{exp}=1.2701(25)$~\cite{Beringer:1900zz}.
The most striking feature of the results in Fig.~\ref{fig:gA-2p} is the
dramatic decrease in $g_A$  as $m_\pi$ decreases from 254 MeV to 202 MeV and on
to 149 MeV. Whereas the $m_\pi\lesssim250\text{ MeV}$ and
$m_\pi\lesssim350\text{ MeV}$ chiral fits for the other observables in
Figs.~\ref{fig:r1v2}--\ref{fig:momfrac}  are consistent, this decrease makes
the two fits  qualitatively different  for  $g_A$.         
In looking for the origin of the strong decrease for $g_A$ at pion masses 202
and 149 MeV we note that when removing excited state contaminants by increasing
$\Delta t$, $g_A$ decreases strongly for both, whereas at 254 MeV it increases
by a comparable amount. 
None of the other observables show this kind of reversal in behavior between
254 MeV and lower pion masses.
Excluding our two lightest $m_\pi$ ensembles, 
our results are consistent with a recent
calculation~\cite{Capitani:2012gj,Jager:2013kha} that uses ${\cal O}(a)$ improved Wilson
fermions and the summation method for $195 \text{ MeV}\lesssim
m_\pi\lesssim649\text{ MeV}$.
In that calculation, chiral extrapolation to the physical mass is
statistically consistent with experiment and $g_A(\Delta t)$ typically
increases 10\% when $\Delta t$ increases by 0.5 fm, which is the behavior for
our 254 MeV ensemble. 
All this evidence suggests that there is an unphysical source of bias that
affects $g_A$  at the small pion masses 202 and 149 MeV more significantly than
the other observables we calculate.

%

\begin{figure}[ht]
  \centering
  \includegraphics[width=.68\textwidth]{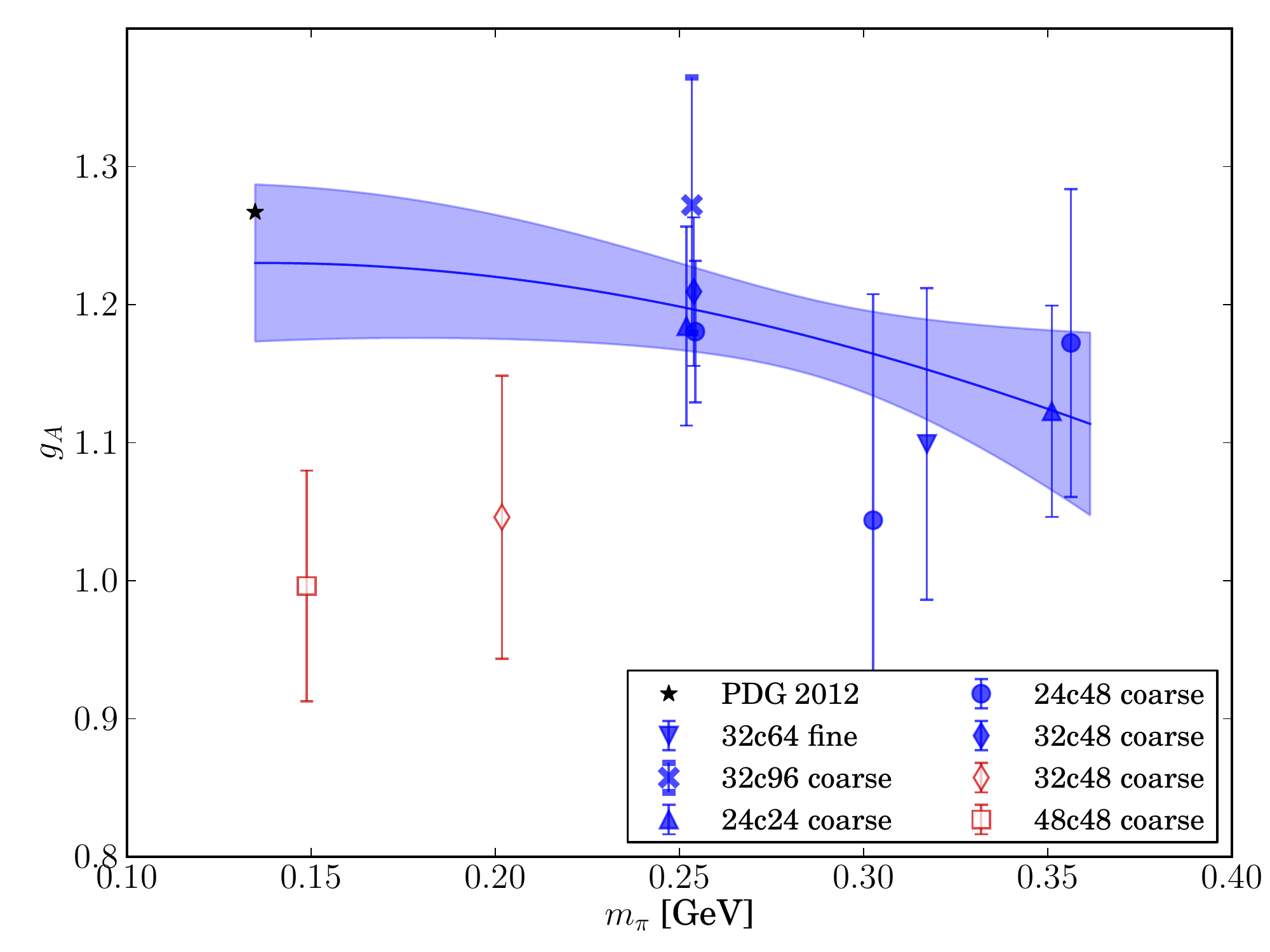} 
  \caption{\label{fig:gt200fig}
    Chiral extrapolation of the axial charge $g_A$ for ensembles with 
    $m_\pi\gtrsim250\text{ MeV}$.}
\end{figure}

One possible source of this bias could be the presence of thermal states.
The lightest state in QCD is a pion at rest, and its contribution to Green's
functions is suppressed as $e^{-m_\pi L_t}$, where $L_t$ is the time extent of
the lattice. 
The ensembles available to us in Tab.~\ref{tab:lat_ens} have values of $m_\pi
L_t$ much smaller than the values $L_t = 2 L_s$ normally used, which may
make our calculations susceptible to thermal effects, especially with lighter
pion masses.
Note that, in contrast to finite-$L_s$ effects (FVE) discussed in the next paragraph,
finite-$L_t$ effects can add contributions to the spectrum that lie below the ground-state
nucleon, significantly changing the time dependence of our two-point and three-point functions
and making it difficult to isolate the desired single-particle hadron state.
Although we find that the behavior of $g_A(\Delta t)$  at $m_\pi\sim250\text{ MeV}$ does not
depend significantly on $m_\pi L_t$,  the qualitative difference between the behavior of
$g_A(\Delta t)$ at $m_\pi=254 \text{ MeV}$ and at 149 and 202 MeV, which both have very small
$m_\pi L_t$, is suggestive of thermal effects for these lower pion masses. 
To assess the effect of the bias in these lowest two masses, we  show the result of a chiral
extrapolation of all the other ensembles in Fig.~\ref{fig:gt200fig}. 
It is noteworthy that this extrapolation is consistent both with experiment and
Ref.~\cite{Capitani:2012gj}, suggesting that the bias is indeed concentrated in  the two
ensembles with small $m_\pi$  and $m_\pi L_t$ that are expected to be particularly susceptible
to thermal state effects.

In other works, discrepancies in $g_A$ have been attributed to finite volume effects (FVE).
According to Ref.~\cite{Yamazaki:2008py}, FVE may lead to a $9\%$ bias in $g_A$
at $m_\pi L_s \approx 4.5$ in calculations with domain wall fermions, and as
much as $\approx 25\%$ in calculations with $N_f=2$ flavors of Wilson fermions. 
However, our data at $m_\pi\approx250\text{ MeV}$ suggest that FVE are
negligible: if we assume that the FVE correction scales as
$\delta^\text{FVE}g_A\sim e^{-m_\pi L_s}$ as in Ref.~\cite{Yamazaki:2008py}, and
use the $g_A$ values from the two lattices with $m_\pi\approx250\text{ MeV}$
that differ only in spatial volume ($m_\pi L_s=3.6$ and $4.8$), we obtain the
estimate $\delta^\text{FVE}g_A=-0.02(6)$ for $m_\pi=149\text{ MeV}$,
$L_s=5.6\text{ fm}$, which is much smaller than the observed discrepancy
$\delta g_A \sim - 0.30$.

\section{Discussion}

The crucial advances in this work are the careful and uniform control of excited-state
contamination and extending the pion mass range down to the nearly physical value 
$m_\pi = 149\text{ MeV}$.
Because of this, we have achieved agreement with experiment for the first time
for the Dirac and Pauli radii, the magnetic moment, and the quark momentum
fraction and observed indications of internal inconsistency in  the calculation
for the one observable that disagrees with experiment, the axial charge.
This 
is strong evidence that this combination of methods is
sufficient to correctly estimate and remove systematic uncertainties in most
nucleon structure calculations and diagnose cases in which some additional bias
renders the calculation unreliable.

Although the lightest pion mass used in our calculation is still above the
physical one, only a very short-range extrapolation is required.
As Figs.~\ref{fig:r1v2}--\ref{fig:momfrac} display, the extrapolated results
are strongly influenced by the lowest pion mass data, which are themselves
already  within $1\sigma$ of experiment for three observables and nearly so
for $(r_1^2 )^v $.
Moreover, the change in the values of our fits from the lowest pion mass,
$m_{\pi } = 149\text{ MeV}$, to the physical point, 
$m_{\pi } = 134.8\text{ MeV}$,
is smaller than, or in the case of $(r_1^2 )^v $ comparable to, the
statistical uncertainty of the $m_{\pi } = 149\text{ MeV}$ lattice data.
Thus,  in contrast to all previous calculations, the dominant uncertainties in
the final results are the statistical uncertainties in the  lowest mass data
points rather than the extrapolation in $m_{\pi }$.

Although additional statistics are necessary to reduce the statistical 
uncertainties of the results to match experimental ones, 
the current precision is a major advance. 
We have achieved $5\%$ precision for the Dirac radius $(r_1^2)^v$, 
$10\%$ precision for the anomalous magnetic moment $\kappa_v$ and the Pauli
radius $\kappa^v(r_2^2)^v$, and $15\%$ precision for the isovector quark
momentum fraction $\langle x\rangle_{u-d}$ (see Tab.~\ref{tab:fits}).
Remarkably, the precision of $(r_1^2)^v$ is already comparable to the
discrepancy between the two contradictory experimental values, and future
improved calculations will contribute to a resolution of this puzzle.

It is useful to comment on the role of ChPT in this work. 
As emphasized above,  because of the small extrapolation distance 
to the physical pion mass,  our primary results  for observables 
are independent of ChPT and its range of convergence. 
However, our data provide interesting input to the study of it. 
We use standard parameters from ChPT phenomenology  given in 
Tab.~\ref{tab:fits} and do not attempt a global fit  to determine them.
When fitting data in the two ranges  $m_\pi\lesssim250\text{ MeV}$  
and $m_\pi\lesssim350\text{ MeV}$, we find consistency for all observables 
except $g_A$. 
In the case of $(r_1^2)^v$, we find that ChPT fits in the two ranges 
become inconsistent if contributions of the $\Delta$ are removed 
(not shown in Fig.~\ref{fig:r1v2}).
Both observations suggest  that ChPT as we are using it is working well 
in this range of pion masses. 
Hence, it is one useful diagnostic for the presence of bias in the calculation 
of $g_A$.

Lattice results are also subject to other systematic errors such as dependence on 
finite lattice spacing $a$ and finite volume. 
However, we expect that these errors are negligible at the present level 
of statistical precision. 
The finite lattice spacing effects, including chiral symmetry breaking, 
should vanish in the limit $a\rightarrow0$. 
Discretization effects of this QCD action in the hadron spectrum have been 
carefully studied in Ref.~\cite{Durr:2010aw} and they were found to be 
quite mild. 
In addition, our results with two values of $a$ near $m_\pi\sim300\text{ MeV}$ 
shown in Figs.~\ref{fig:r1v2}--\ref{fig:momfrac} agree within statistics.

Finite volume effects are widely believed to be negligible if the spatial box
size $L_s\gtrsim 4 m_\pi^{-1}$.  
Most of our ensembles satisfy this criterion,
including the lightest data point $m_\pi=149 \text{ MeV}$. 
Two ensembles at
$m_\pi\approx 250\text{ MeV}$ serve as a direct check that finite volume
effects are insignificant: we have compared data from lattices with $m_\pi
L_s=3.6 \, (24^3 \times 48)$ and $m_\pi L_s=4.8 \, (32^3 \times 48)$, otherwise
completely identical, and find no statistically significant discrepancies in
any of the reported quantities.

In the case of $g_A$, we have noted  that internal inconsistencies  such as
the lack of agreement between the two ranges of chiral fits 
and the qualitative difference between the behavior of $g_A(\Delta t)$ 
at $m_\pi=254 \text{ MeV}$ and at lower pion masses indicate some source 
of bias  that is not present for the other observables. 
One possible  source of bias could be  the presence of thermal states,  
as discussed  in connection with Fig.~\ref{fig:gt200fig}.
Another possibility is the breaking of chiral symmetry by Wilson fermions. 
Although we have verified that finite $a$ effects are smaller than 
the statistical uncertainty at  $m_\pi= 300\text{ MeV}$, it is possible that
chiral symmetry  breaking leads to a progressively stronger systematic bias 
at a given $a$ as the pion mass is lowered, and that this disproportionately
affects the axial charge.
Yet another possibility is very long-range autocorrelations as recently reported 
for a calculation~\cite{Ohta:2013qda} of $g_A $ using domain wall fermions
with $m_\pi$ down to $170\text{ MeV}$, 
where values for $g_A $  averaged over the first and second halves  
of the ensemble differed by almost four standard deviations while 
no other observables showed similar  autocorrelations.
While the physical  cause has not yet been determined, slow equilibration
across topological sectors  could conceivably affect the axial charge.
Although we do not observe such anomalies in our calculation, 
they may still lead to underestimated stochastic error of $g_A$ and apparent
deviation from experiment.
We also note that a recent preliminary report on work with twisted mass
fermions using a $48^3\times96$ ensemble at the physical pion mass~\cite{Alexandrou:2013jsa}
showed agreement with the experimental value for $g_A$. 
Although it had no study or removal of excited–state effects, 
a small volume corresponding to $m_\pi L = 3.0$, $N_f=2$, and included no data 
on electromagnetic form factors or radii, we expect that further results 
from that work will be useful in helping to resolve the problem with $g_A$.

More lattice spacings and further volumes and temporal extents at pion masses
below $200\text{ MeV}$ are required to obtain strong bounds on the
systematic uncertainties of our calculations, and in particular to
understand the unique behavior of the axial charge. With appropriately
increased statistics, our control over excited states could be
cross-checked by further varying the nucleon source-sink separation and
by comparing against other excited state removal techniques.

It is a significant advance that Lattice QCD, for four key properties of
nucleon structure, now is in good agreement with experimental results.

\section*{Acknowledgements}
We thank Zoltan Fodor for useful discussions and the Budapest-Marseille-Wuppertal collaboration
for making some of their configurations available to us. 
This research used resources of the Argonne Leadership Computing Facility at 
Argonne National Laboratory, which is supported by the Office of Science 
of the U.S. Department of Energy under contract \#DE--AC02--06CH11357, 
and resources at Forschungszentrum J\"ulich. 
During this research JRG, SK, JWN, AVP and SNS were supported in part 
by the U.S. Department of Energy Office of Nuclear Physics under grant \#DE--FG02--94ER40818, 
ME was supported by DOE grant \#DE--FG02--96ER40965, 
SNS was supported by the Office of Nuclear Physics in the US Department of Energy's 
Office of Science under Contract \#DE--AC02--05CH11231 
and by RIKEN Foreign Postdoctoral Researcher Program,
and SK was supported in part by Deutsche Forschungsgemeinschaft through grant SFB--TRR 55.

\bibliographystyle{elsarticle-num}
\bibliography{text}

\end{document}